\documentclass[%
 aip,
 amsmath,amssymb,
 reprint,%
]{revtex4-1}

\usepackage{graphicx}
\usepackage{dcolumn}
\usepackage{bm}
\usepackage{newtxtext,newtxmath}
\usepackage{graphicx}
\usepackage[dvipsnames]{xcolor}
\usepackage[letterpaper,margin=1in]{geometry}
\usepackage[utf8]{inputenc}
\usepackage[T1]{fontenc}

\usepackage{url}
\usepackage{etoolbox}

\makeatletter
\def\@email#1#2{%
 \endgroup
 \patchcmd{\titleblock@produce}
  {\frontmatter@RRAPformat}
  {\frontmatter@RRAPformat{\produce@RRAP{*#1\href{mailto:#2}{#2}}}\frontmatter@RRAPformat}
  {}{}
}%
\makeatother
\begin{document}

\preprint{AIP/123-QED}

\title{Comparison and optimisation of hybridization algorithms for onboard classical and quantum accelerometers}
\author{Benoit Kaczmarczuk}

\affiliation{ 
Laboratoire Temps Espace, Observatoire de Paris, Université PSL, Sorbonne Université, Université de Lille, LNE, CNRS, 61 avenue de l'Observatoire, 75014 Paris, France
}%

\author{Yannick Bidel}
\author{Alexandre Bresson}
\author{Nassim Zahzam}
\author{Alexis Bonnin}
\author{Malo Cadoret}
\affiliation{%
DPHY, ONERA, Université Paris Saclay, 91123 Palaiseau, France
}%

\author{Tim Enzlberger Jensen}
\affiliation{%
National Space Institute, Technical University of Denmark
}%
\author{Quentin Beaufils}%
\author{Franck Pereira dos Santos}%
\affiliation{ 
Laboratoire Temps Espace, Observatoire de Paris, Université PSL, Sorbonne Université, Université de Lille, LNE, CNRS, 61 avenue de l'Observatoire, 75014 Paris, France
}%

\date{\today}

\begin{abstract}
We study two hybridization algorithms used for the combination of a quantum inertial sensor based on atom interferometry with a classical inertial sensor for onboard acceleration measurements. The first is based on the direct extraction of the interferometer phase, and was previously used in seaborne and airborne gravity measurement campaigns. The second is based on the combination of three consecutive measurements and was originally developed to increase the measurement range of the quantum sensor beyond its linear range. After comparing their performances using synthetic data, we implement them on acceleration data collected in a recent airborne campaign and evaluate the bias and the scale factor error of the classical sensor. We then extend their scope to the dynamical evaluation of other key  measurement parameters (e.g. alignment errors). We demonstrate an improvement in the correlation between the two accelerometers' measurements and a significant reduction of the error in the estimation of the bias of the classical sensor.
\end{abstract}

\maketitle

\section{Introduction}

Quantum sensors based on atom interferometry have demonstrated their ability to perform absolute and long term stable measurements of inertial quantities \cite{Geiger2020}, owing to the stability of their scale factor and to the good control of their systematics. Yet, they suffer from measurement dead times and aliasing\cite{Joyet2012} of inertial noise, which limit their range of applications, in particular in the domain of inertial navigation. Moreover, their non-linear response, related to the cosine of the interferometer phase, leads to ambiguities in the determination of the inertial quantities, when the amplitude of their variations exceeds the range of one interferometer fringe. Methods have been developed to overcome these issues, such as realizing joint measurements for no dead-time operation \cite{Dutta2016,Savoie2018}, and hybridizing quantum and classical sensors \cite{Lautier2014}. This allowed in particular for operating the sensors onboard ships \cite{Bidel2018,Zhou2024,Qiao2025} and planes \cite{Geiger2011,Bidel2020,Bidel2023}, in the presence of large fluctuations of acceleration, and would allow improving accelerometer measurements onboard satellites for geodesy applications \cite{Abrykosov2019,Zahzam2022,Hosseiniarani2025}.

In this context, optimizing the hybridization algorithms is necessary to take the full benefit of the combination of the two sensors, and deliver continuous bias free measurements with the best possible fidelity. Various methods have been demonstrated, based on direct phase extraction \cite{Bidel2018}, mid-fringe lock \cite{Lautier2014} or Kalman filtering \cite{Cheiney2019,Hosseiniarani2025,Huang2023}.
In this paper, we compare two different hybridization algorithms, a first one (Algo I) developed by the ONERA team for their onboard campaigns for correcting the classical sensor from its bias and scale factor fluctuations \cite{Bidel2018}, and a second one (Algo II) originally developed by LTE for extending the range of operation of a quantum gravimeter \cite{Merlet2009} that we adapted for this study. We compare their performance and merits using both real and simulated data. Additionally, we show how they can be adapted to continuously estimate other key measurement parameters, such as axis cross couplings and accelerometer transfer function parameters, and track their variation rather than fixing them to predetermined values. Finally, we show that they allow for coping with rotation noise with the help of additional measurements by rotation sensors, which opens perspectives for performing high sensitivity strapdown measurements.

\section{Description of the hybridization algorithms}

\subsection{Context}

We consider a Quantum Accelerometer (QA) consisting of a cold atom interferometer realized with a standard sequence of three Raman pulses, similar to the one described in \cite{Geiger2020}. The measured quantity at the output of the interferometer is the transition probability between the two $^{87}$Rb hyperfine levels of the ground state:

\begin{equation}
P_i=P_0-\frac{C_i}{2}\cos(\Phi_i) + \delta P_i
\label{eq}\end{equation}

where $P_0$ is the offset of the interference fringes, $C_i$ the contrast and $\delta P_i$ the contribution related to detection noise, which we will consider where applicable as a white noise with standard deviation $\sigma_P$. Here the index $i$ enumerates the successive measurements. The phase difference accumulated between the two interferometer paths $\Phi_i$ is proportional to the acceleration along the measurement axis as well as rotation induced inertial pseudo-accelerations (Coriolis, centrifugal and Euler). Here, we neglect other phase contributions related to non-inertial effects (such as due to light shifts, magnetic field gradient, wavefront distortions ...). The linear acceleration contribution to the phase is given by $\Phi=k_{eff}aT^2$, where $k_{eff}$ is the Raman effective wave-vector, $a$ the acceleration, and $T$ is the free separation time between two consecutive Raman pulses. In practice, a frequency chirp $\alpha$ can be applied to the frequency difference between the Raman lasers to keep them on resonance, which adds a contribution $\alpha T^2$, so that the total phase is finally given by $\Phi=k_{eff} a T^2-\alpha T^2$. Note that the expressions above, valid for a constant acceleration, can be generalized to fluctuating accelerations, by replacing $a$ by its weighted average over the interferometer duration, using as a weighting function the triangular-shape acceleration transfer function $g_a$ \cite{Varoquaux2009}. 

For large acceleration fluctuations, the variation of the phase between successive measurements can exceed $2\pi$, causing an ambiguity when deriving this phase from the corresponding transition probability, which can be lifted by hybridizing the QA with a classical accelerometer (CA). The hybridization strategy consists in combining measurements of the QA, which delivers absolute measurements with a low bandwidth and dead times to the continuous ones of a classical accelerometer CA, which has higher data rate and bandwidth, but suffers from bias instability, so as to obtain hybrid measurements which combine the best features of the two technologies. In practice, the two algorithms studied here use measurements for the QA to obtain the best estimate of the bias and scale factor of the CA.

\subsection{Algo I}

 The first hybridization algorithm (Algo I) uses a method based on the direct extraction of the phase from the transition probability measurement \cite{Bidel2018}. 
 
Neglecting detection noise, the transition probability measured at the i-th cycle writes as:
\begin{equation}
\begin{split}
    P_i&=P_0- \frac{C}{2} \cos \left(k_{eff} T^2 a_{QA,i} +\phi_{control,i}\right)\\
\end{split}
\label{eq:2}
\end{equation}
where $a_{QA,i}$ is the acceleration variations experienced by the QA around an offset value set close to $g$ 

using $\phi_{control,i}$, an additional controlled phase that can be applied to the QA phase (for example by chirping the frequency of the interrogation lasers). 
 
We write the relation between $a_{QA,i}$ and the acceleration measurement given by the CA $a_{CA,i}$ as:

\begin{equation}
a_{QA,i}=\eta a_{CA,i}+b
\end{equation}

with undetermined bias $b$ and scale factor $\eta$. Here, the high-rate data from the CA are integrated with the triangular shaped weighting function of the QA to obtain comparable measurements at the same rate. 

At each measurement cycle $i$, the method below provides an iterative  estimate of the bias $\hat{b}_i$ and the scale factor $\hat{\eta}_i$ of the CA, which allows to derive $\hat{a}_{c,i}$, a best estimate of the acceleration experienced by the CA by correcting its measurements as : $\hat{a}_{c,i} = \hat{\eta}_i a_{CA,i} +\hat{b}_i$. 
 
When trying to extract the acceleration measured by the QA out of the measured transition probability, possible solutions are $\hat{a}_{q,i}=(\pm \arccos(2(P_0-P_i)/C)-\phi_{control,i}+2k\pi)/k_{eff} T^2$, with $k$ an integer. $\hat{a}_{c,i}$ is used to lift the ambiguity: the chosen solution is the one that is the closest to $\hat{a}_{c,i}$. The difference $\hat{a}_{q,i}-\hat{a}_{c,i}$ is then used to iteratively correct the estimate of the bias \cite{Salducci2025} as :  

\begin{equation}
    \hat{b}_{i+1}=\hat{b}_{i}+G_b(\hat{a}_{q,i}-\hat{a}_{c,i})
\end{equation}

and the estimate of the scale factor as:

\begin{equation}
\hat{\eta}_{i+1}=\hat{\eta}_{i}+G_{\eta}(\hat{a}_{q,i}-\hat{a}_{c,i})/a_{CA,i}
\label{eq:algoIeta}
\end{equation}

where $G_b$ and $G_{\eta}$ are the gains of the integrator loops. Since the denominator $a_{CA,i}$ can get arbitrarily close to zero, we regularize it to a pseudo-inverse to avoid divergence in the applied corrections. 
$\sigma^2$ is estimated with exponential averages of $a_{CA,i}$ with a characteristic time of $10$ s . 

\begin{equation}
\hat{\eta}_{i+1}=\hat{\eta}_{i}+G_{\eta}(\hat{a}_{q,i}-\hat{a}_{c,i})\times a_{CA,i} /(a_{CA,i}^2+\sigma^2).
\label{eq:4}
\end{equation}

Note that it is also possible to regularize by replacing  $1/a_{CA,i}$ with $a_{CA,i}$ in equation (\ref{eq:algoIeta}) \cite{Bidel2018,Bidel2020}. 

\subsection{Algo II}

The second algorithm (Algo II) is an adaptation of the non-linear fringe tracking lock described in \cite{Merlet2009}. It allows to determine the bias and the scale factor of the CA by correcting their values using three consecutive joint measurements of the quantum and classical sensors. 

Like Algo I, we mark the algorithm estimates of physical quantities (e.g. $b$) with a circumflex ($\hat{b}$). 
Following \cite{Merlet2009}, the estimate of the CA bias  $\hat{b}_i$ can be corrected as:
\begin{equation}
    \hat{b}_{i+1}=\hat{b}_i + G_b^{'} \times \frac{1}{k_{eff} T^2} \frac{N_i}{D_i}  .
\end{equation}

where we have introduced the intermediate quantities $N_i$ and $D_i$:

\begin{align}
    N_i = (P_{i-2}-P_{i-1})(\cos \hat{\phi}_i-\cos \hat{\phi}_{i-1}) \nonumber\\
    -(P_i-P_{i-1} )(\cos \hat{\phi}_{i-2}- \cos \hat{\phi}_{i-1}) 
\end{align}

\begin{align}
    D_i=(\sin \hat{\phi}_{i-2}-\sin \hat{\phi}_{i-1} )(\cos \hat{\phi}_i- \cos \hat{\phi}_{i-1})\nonumber \\
    -(\sin \hat{\phi}_i-\sin \hat{\phi}_{i-1} )(\cos \hat{\phi}_{i-2}- \cos \hat{\phi}_{i-1}) 
\end{align}

where $\hat{\phi}_i=k_{eff} T^2 (\hat{\eta}_i a_{CA,i}+\hat{b}_i)+\phi_{control,i}$ is the estimate of the interferometer phase at the i-th measurement. 

As for Algo I, since $D_i$ can get close to 0, we replace it with $(D_i^2 + \sigma_D^2)/D_i$ where $\sigma_D$ is the standard deviation of $D_i$ :
\begin{equation}
    \hat{b}_{i+1} = \hat{b}_i + G_b^{'} \times \frac{1}{k T^2} N_i \frac{D_i}{D_i^2 + \sigma_D^2 }
\end{equation}

The estimate of the CA scale factor can then be corrected at the i-th interferometric cycle as: 

\begin{equation}
   \hat{\eta}_{i+1} = \hat{\eta}_i + G_{\eta}^{'} \times \frac{1}{k T^2} N_i \frac{D'_i}{(D'_i)^2 + \sigma_{D}'^2}
   \label{eq:5}
\end{equation}
with 
\begin{align}
    D'_i&=
    (\sin \hat{\phi}_{i-2}-\sin \hat{\phi}_{i-1} )\nonumber\\
    &\times (a_{CA,i} \cos \hat{\phi}_i - a_{CA,i-1}\cos \hat{\phi}_{i-1})\nonumber\\
    &- (\sin \hat{\phi}_i-\sin \hat{\phi}_{i-1} ) \nonumber\\
    & \times(a_{CA,i-2} \cos \hat{\phi}_{i-2}- a_{CA,i-1}\cos \hat{\phi}_{i-1}) 
\end{align}

$\sigma_D^2$ and $\sigma_D'^2$ are estimated with exponential averages of $D_i^2$ and $(D_i')^2$ with a characteristic time of $10$ s .

In addition, since significant variations of the interferometer phase are required for this algorithm to operate, modulating the applied phase $\phi_{control,i}$, for instance with a three-step modulation $[\pi/2;0;-\pi/2]$ as in \cite{Merlet2009} or randomly over $2\pi$, allows the method to remain efficient whatever the level of acceleration fluctuations. Note that Algo II eliminates the need for an independent estimate of the contrast $C$ and average probability $P_0$, unlike Algo I.

\section{Tests and comparisons}

\subsection{With synthetic data}

To compare the performance of the two hybridization algorithms, we first implement them on a common dataset generated synthetically. The QA accelerations $a_{QA}$ are randomly drawn in a normal distribution with $0.38$ m/s$^2$ standard deviation. To account for the imperfect correlation between the two accelerometers and errors in the knowledge of the CA scale factor and bias, we set $a_{CA}=(a_{QA}-b)/\eta +\delta_a$, with $\delta_a$ an uncorrelated part of the CA signal drawn in a normal distribution with a standard deviation $\sigma_{\delta_a}=4.8\times 10^{-5}~\text{m}~\text{s}^{-2}$, $\eta = 1.001$ and $b=2 \times 10^{-5}~\text{m}~\text{s}^{-2}$. Detection noise, with standard deviation $\sigma_P=0.016$, is added to the calculated transition probabilities, for the interferometer duration $2T=40 \text{ms}$. The cycle time is taken to be $T_c=0.1$ s.

\begin{figure}[h]
    \centering
    \includegraphics[width=0.8\linewidth]{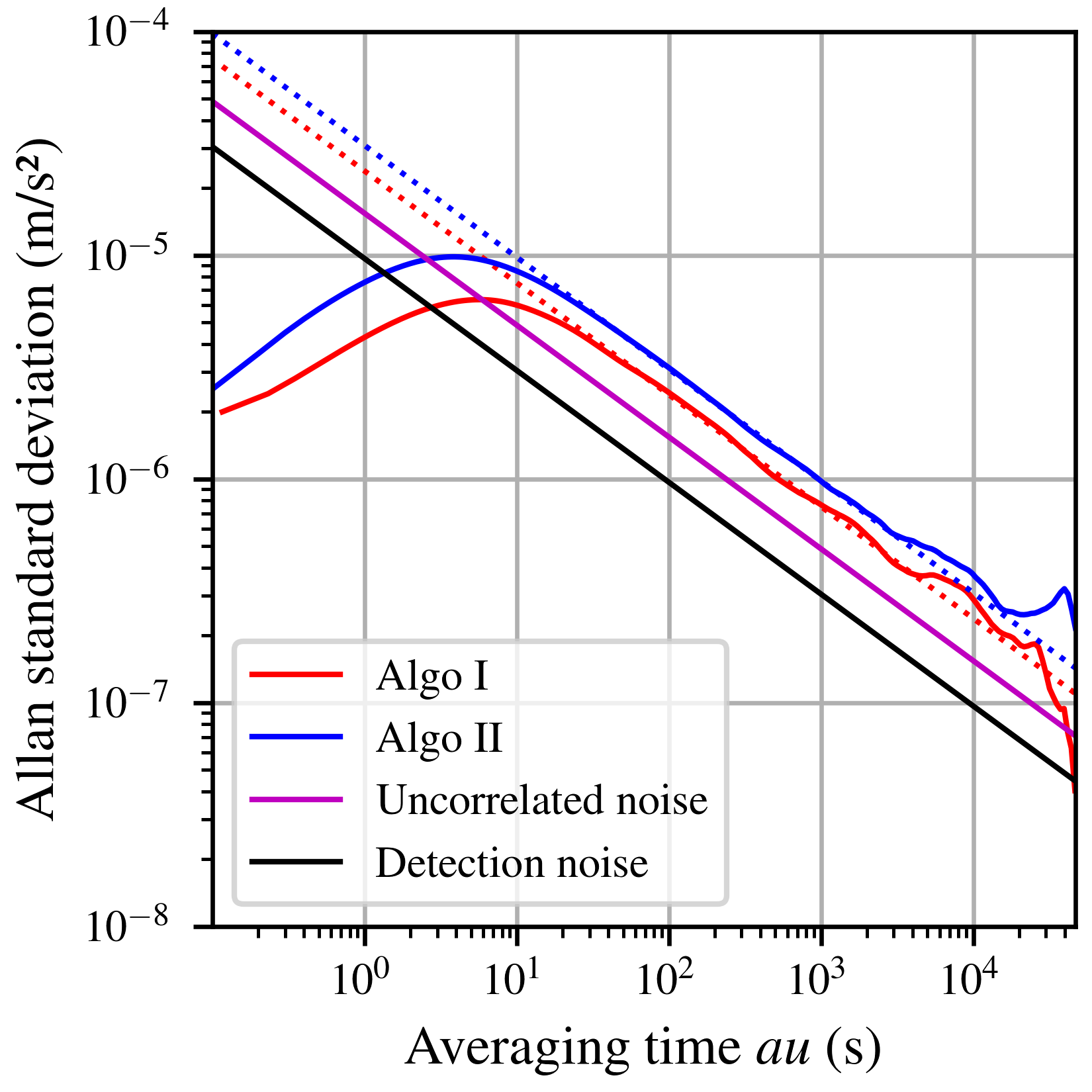}
    \caption{The Allan standard deviations of bias estimates for Algo I (in red) and Algo II (in blue).}
    \label{fig:Figure01}
\end{figure}

We then implement both algorithms on the synthetic datasets to determine estimates of the scale factor $\eta$ and of the bias $b$, adjusting the gains for equating the time constants (0.2 for Algo I and 0.24 for Algo II), for both quantities and both algorithms. Fig. \ref{fig:Figure01} displays the stability of the bias estimates for both algorithms, characterized by their Allan standard deviations \cite{Allan1972,Allan1971}, assuming the contrast $C$ and the transition probability $P_0$ are known. For this simulation, we choose $C=0.23$ and $P_0=0.5$. For averaging times larger than the time constant of the loop, both stabilities decrease as $1/\tau^{1/2}$, with $\tau$ the averaging time, with extrapolated values at 1 cycle given by $7.0\times10^{-5}~\text{m}~\text{s}^{-2}$ (Algo I) and $9.8\times10^{-5}~\text{m}~\text{s}^{-2}$ (Algo II). For comparison, we display as solid lines the Allan standard deviations of the fluctuations induced by the uncorrelated acceleration noise ($4.9\times10^{-5}~\text{m}~\text{s}^{-2}$  at 1 cycle) and by the detection noise ($3.0\times10^{-5}~\text{m}~\text{s}^{-2}$ at 1 cycle). The stabilities of bias estimates lie above the quadratic sum of these two contributions ($5.75\times10^{-5}~\text{m}~\text{s}^{-2}$ 
 at 1 cycle), by about 20 \% for Algo I, and by about 70 \% for Algo II owing to the larger number of successive measurements needed for estimating the difference between the accelerations measured by the quantum and classical sensors.

\begin{figure}[h]
    \centering
    \includegraphics[width=1\linewidth]{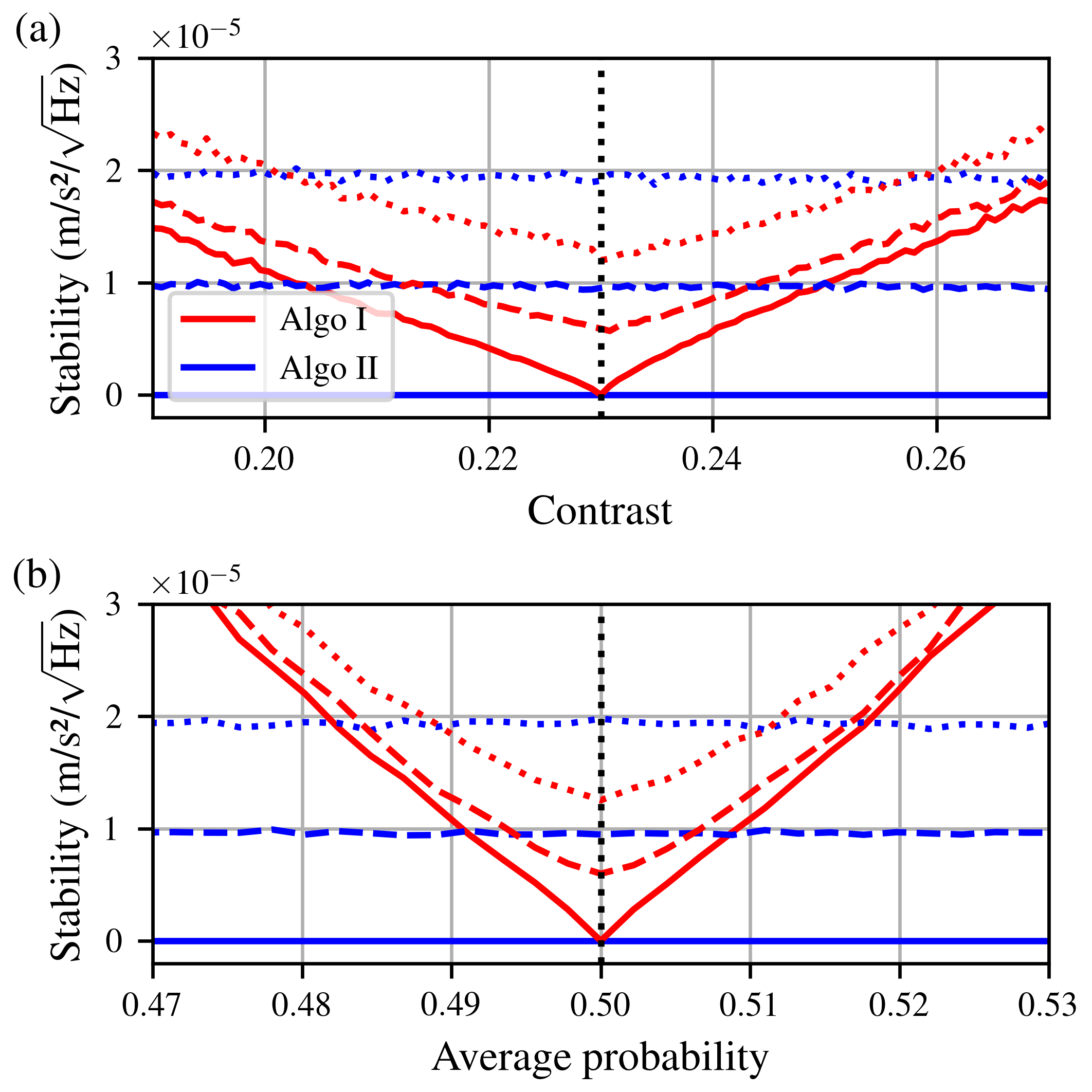}
    \caption{Impact of errors in the contrast $C$ and in the offset $P_0$ used in the algorithms on the stability of bias estimates with no detection noise. The standard deviation of the uncorrelated acceleration noise $\sigma_{\delta_a}$ is 0 (solid line), $1.8\times 10^{-5}~\text{m} \text{s}^2$ (dotted line) and $3.6\times 10^{-5}~\text{m} \text{s}^2$ (dashed line)}. 
    \label{fig:Figure02}
\end{figure}

We then compare the algorithms in the case of imperfect knowledge of the contrast $C$ (Fig. \ref{fig:Figure02}. (a)) and mean transition probability $P_{0}$ (Fig. \ref{fig:Figure02}. (b)), using as a figure of merit for both algorithms the stability  of the bias estimates (expressed in m/s$^2$/Hz$^{1/2}$). This allows to assess their robustness with respect to errors in the knowledge of one of these parameters, taking the other one as perfectly known. While we observe that Algo I performs sligthly better than Algo II in the absence of errors, Algo I degrades with increasing errors, while Algo II is not affected. Eventually, for large errors, Algo II performs better. This comes from the fact that, for Algo I, errors in the value of $C$ and $P_{0}$ used as input parameters in the algorithm result in errors on the extracted phase. Besides, since the direct phase extraction lead in some cases to undefined mathematical solutions, one has to drop out the corresponding measurement points, whose fraction increases with the amplitude of the detection noise. In contrast, Algo II does not suffer from this drawback. It also completely eliminates the need to use an input estimate for $P_0$ while an error in $C$ only affects the gain of the loop.

\begin{figure}[h]
    \centering
    \includegraphics[width=0.8\linewidth]{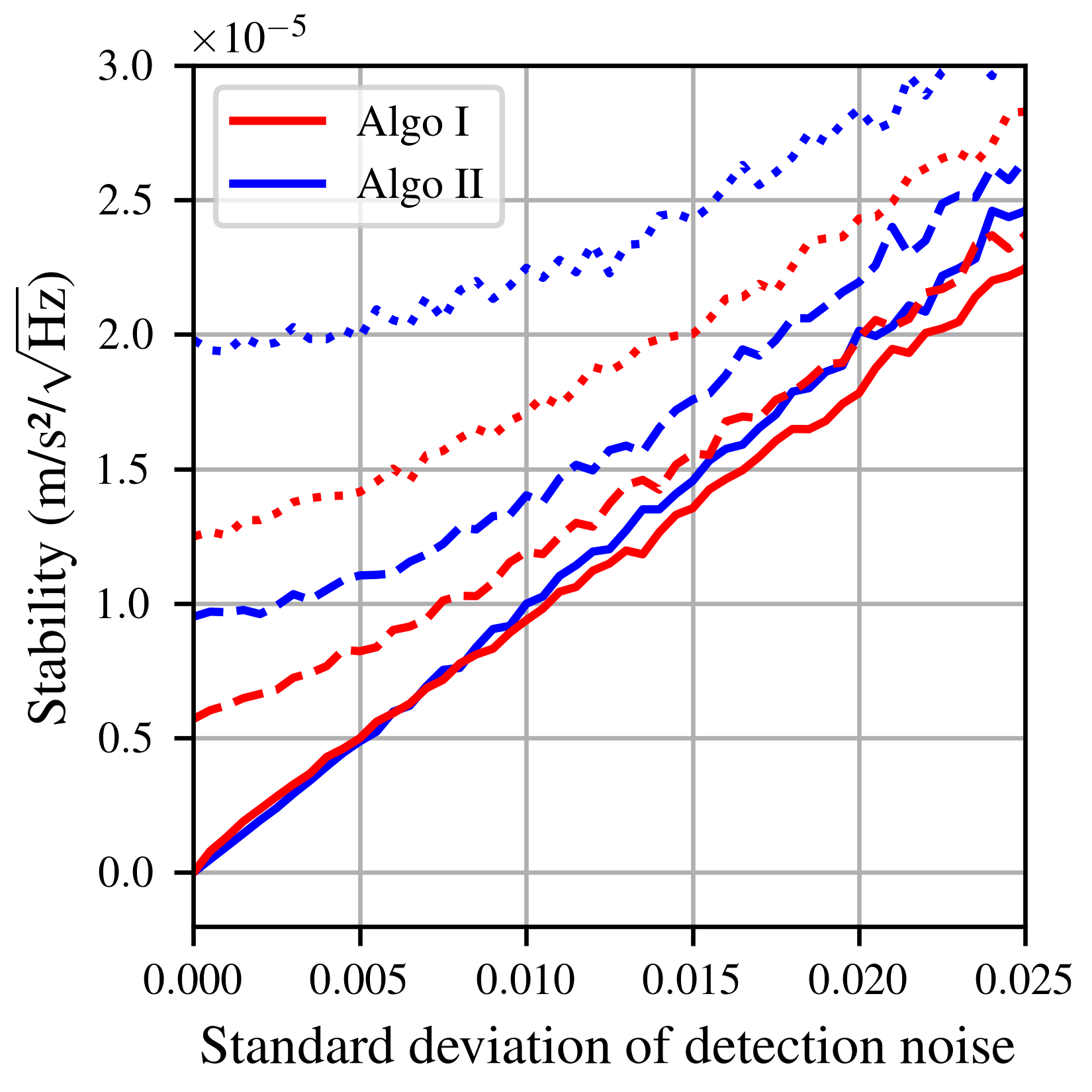}
    \caption{Impact of the detection noise on the stability of the bias estimates. The standard deviation of the uncorrelated acceleration noise $\sigma_{\delta_a}$ is 0 (solid line), $1.8\times 10^{-5}~\text{m} \text{s}^2$ (dotted line) and $3.6\times 10^{-5}~\text{m}$.
    }.
    \label{fig:Figure03}
\end{figure}

Fig. \ref{fig:Figure03} displays the stability of the bias estimates as a function of the amplitude of detection noise, for different amplitudes of uncorrelated acceleration noise, and for the two algorithms. In the presence of uncorrelated acceleration noise, the obtained curves follow as expected a quadratic behaviour, since fluctuations induced by detection noise and uncorrelated accelerations are independent and add up quadratically. Both algorithms lead to similar stabilities in the absence of uncorrelated noise, but Algo I behaves better than Algo II in their presence. 

Finally, we point out that $C$ and $P_0$ can be estimated from of the data, for instance by calculating moments of the transition probability measurements or fitting the density distribution of the transition probability \cite{Geiger2011,Kaczmarczuk2025}. This can reduce the impact of errors in their knowledge on the use of Algo I. 

\subsection{With real data}

We applied both algorithms to data obtained during an airborne measurement campaign over Greenland with the absolute quantum gravimeter GIRAFE \cite{Jensen2025}. For these measurements, the duration of the interferometer was set to $2T=40$ms. We set the value of the lock gains to $0.03$ for algo I and $0.036$ for algo II in order to reach the same convergence time ($\sim 5$ s).  

\begin{figure}[h]
    \centering
    \includegraphics[width=1\linewidth]{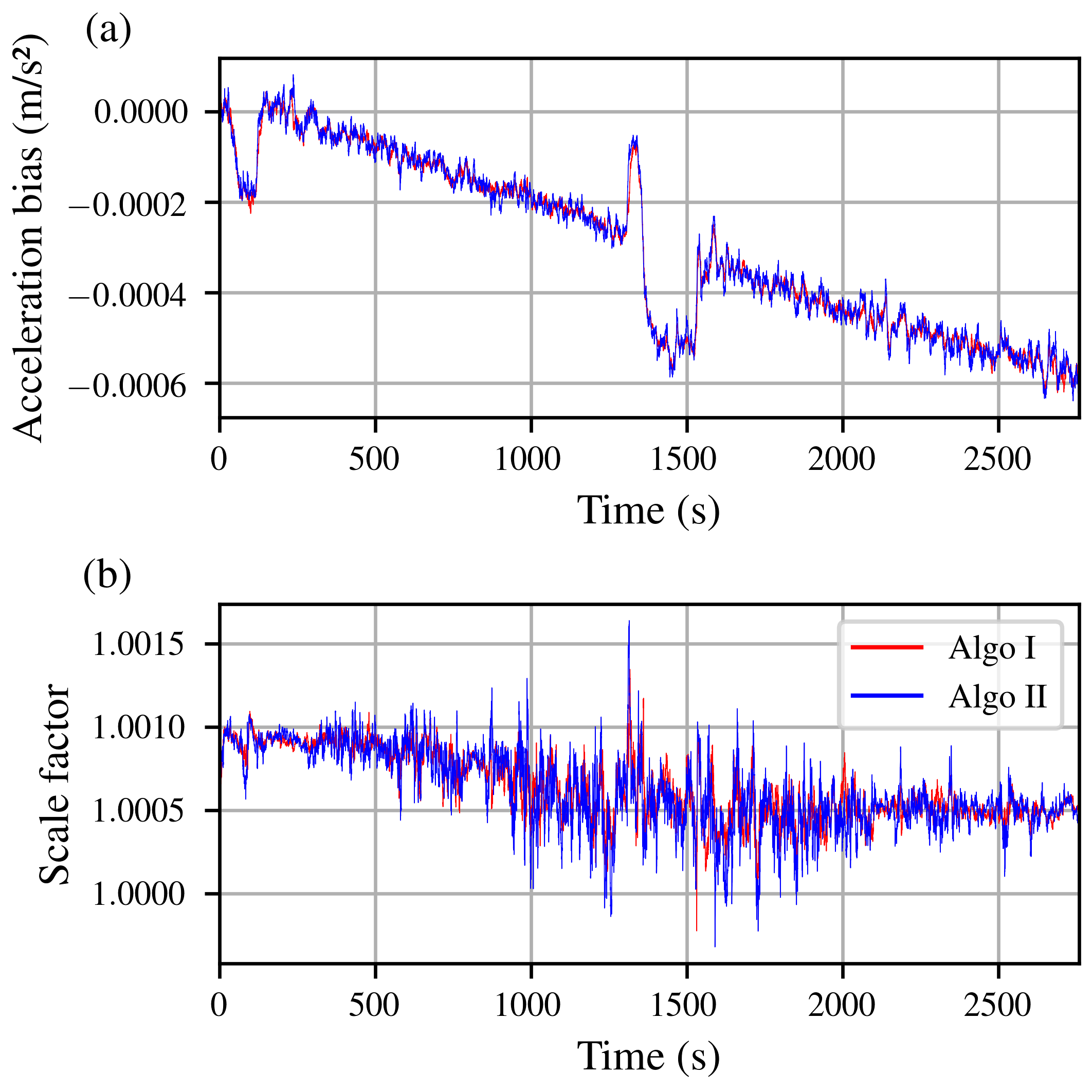}
    \caption{Extracted bias (a) and relative scale factor (b) corrections of the classical accelerometer, for the two different hybridization algorithms.}
    \label{fig:Figure04}
\end{figure}

With real measurements, new effects appeared that were not included in the simulation. First, the CA is composed of three accelerometers mounted in a trihedron, with its vertical measurement axis potentially not perfectly aligned with the measurement axis of the atomic accelerometer, which induces axis-crossing errors. Secondly, the classical sensor acts as a filter, which limits the quality of the correlation with the quantum measurement. Its transfer function is modeled as a second order low pass filter, and we apply corrections to compensate for it. 

In practice, the acceleration given by the CA is now written : 

\begin{equation}
\hat{a}_{c,i} = \hat{\eta}_i a_{z,i}+\eta' a'_{z,i} +\eta'' a''_{z,i}+\eta_x a_{x,i}+\eta_y a_{y,i} +\hat{b}_i
\end{equation}

where $a_{z,i}$ is the acceleration along the z-axis of interest, $a_{x,i}$ and $a_{y,i}$ are the accelerations measured along the transverse axes, and $\eta_x$ and $\eta_y$ are sensitivity coefficients allowing to account for misalignments or cross-axes couplings. To account for the CA's transfer function,  $a'_{z,i}$ and  $a''_{z,i}$ are obtained by the convolution the CA's acceleration data with respectively the first and second derivative of the triangular shaped transfer function of the QA. Here we apply this filter correction only to the measurements along z, neglecting the impact of the transfer functions along the transverse directions. $\eta'$ and $\eta''$ are the sensitivity coefficients related to these contributions $a'_{z,i}$ and  $a''_{z,i}$.

The values of these four sensitivity coefficients are displayed on Table \ref{tab:parametres}. They have been determined prior to the measurement campaign, via a calibration study performed in the lab on a motion simulator.

\begin{table}[h]
    \centering
    \begin{tabular}{|l|c|}
        \hline
        \textbf{Parameter} & \textbf{Value} \\
        \hline
        $\eta'$     & 5.002e-4 \\
        $\eta''$       & 2.068e-7 \\
        $\eta_x$     & -6.5e-4 \\
        $\eta_y$     & 1.95e-3 \\
        \hline
    \end{tabular}
    \caption{Sensitivity coefficients determined via in-lab calibration, realized prior to the onboard operation.}
    \label{tab:parametres}
\end{table}

Fig. \ref{fig:Figure04} displays the evolution of the bias and scale factor of the CA determined by the two algorithms for a round trip of the plane along a line (linear trajectory with a half turn midway). A good agreement between the two methods is found. The bias decreases globally linearly, which we attribute to a linear drift of the temperature during the measurements. However, we observe deviations from the linear trend during aircraft's turning maneuvers at the beginning and midway. These deviations are due to errors in the bias determination arising from dynamical changes of the cross coupling coefficients $\eta_x$ and $\eta_y$ during flight (see section \ref{sectioniv}).

\section{Optimization of the algorithms\label{sectioniv}}

\begin{figure*}
    \centering
    \includegraphics[width=0.85\textwidth]{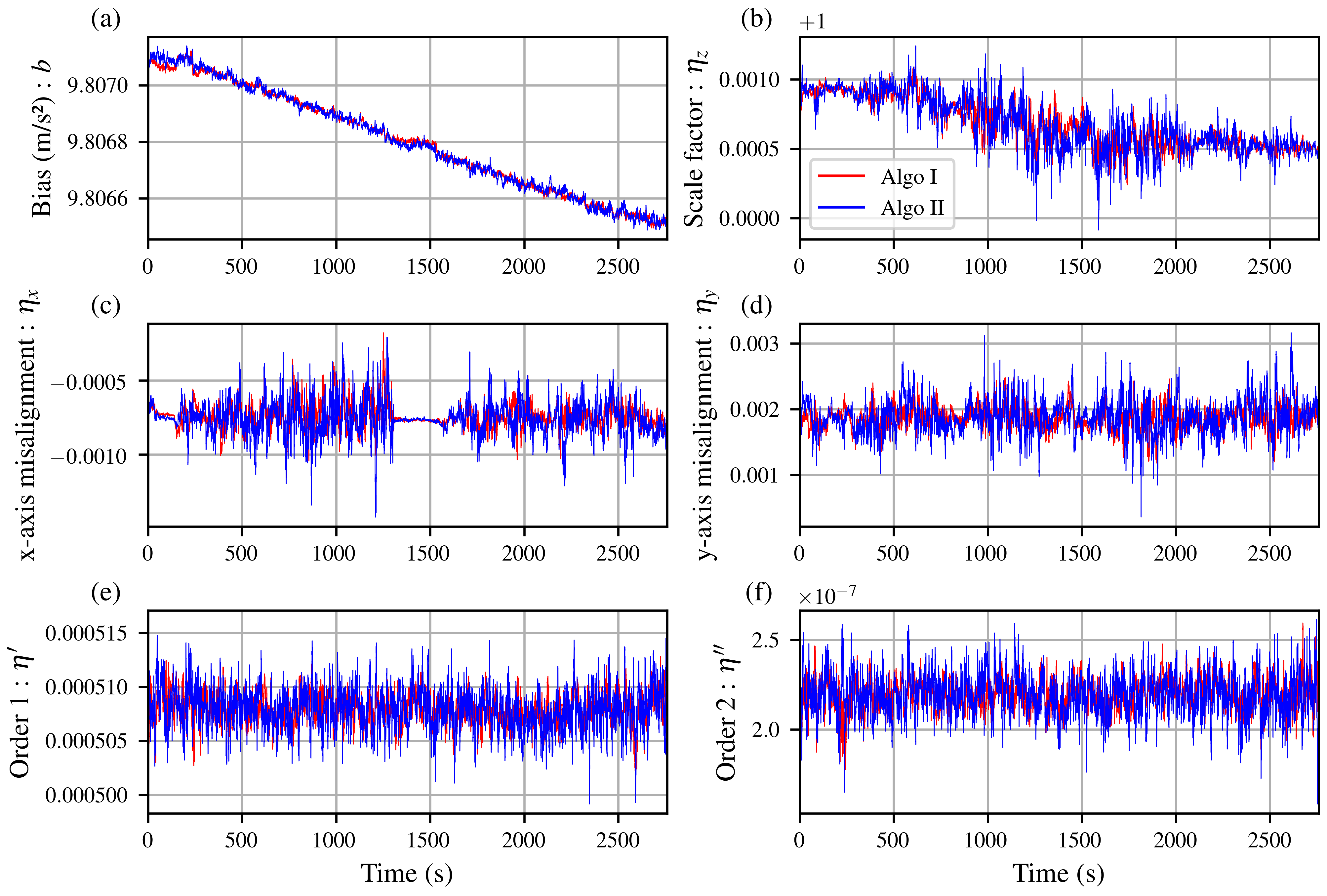}
    \caption{Bias and sensitivity coefficients when all optimized.}
    \label{fig:Figure05}
\end{figure*}

\begin{figure*}
    \centering
    \includegraphics[width=0.8\textwidth]{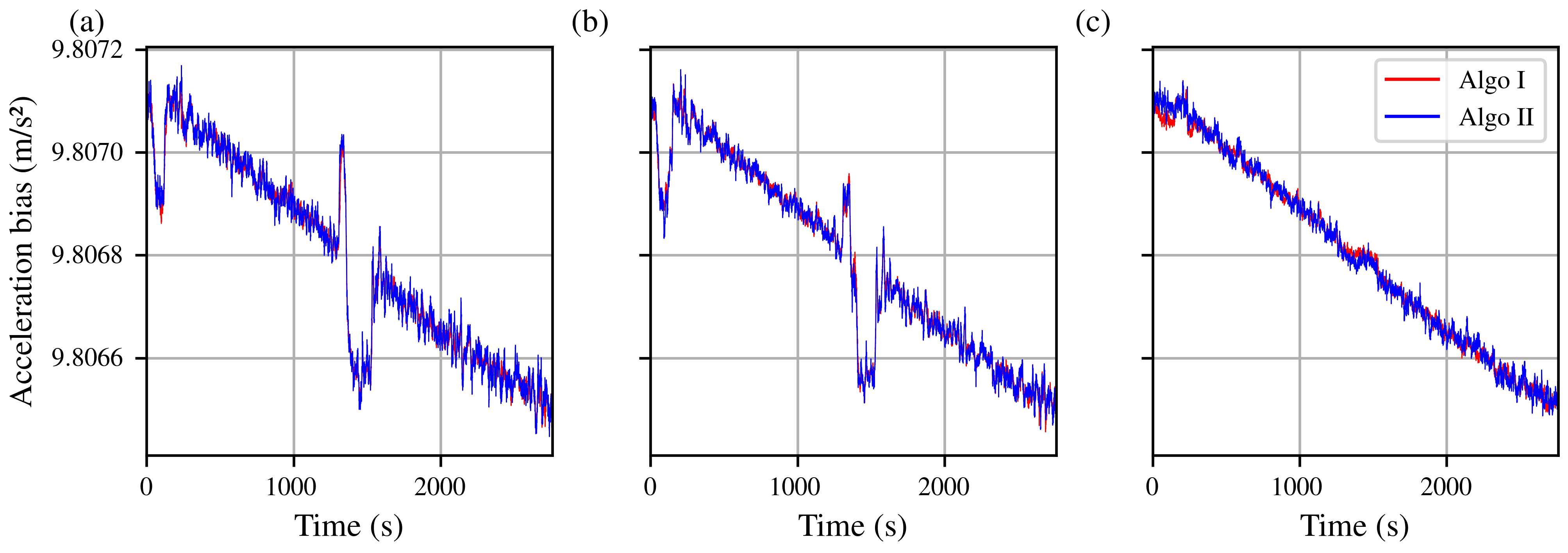}
    \caption{Bias estimated by the different hybridization algorithms : a. Only the scaling factor and bias are optimized. b. All coefficients are optimized except $\eta_x$. c. All coefficients are optimized.}
    \label{fig:Figure06}
\end{figure*}

Both algorithms can be adapted to estimate dynamically the four sensitivity coefficients discussed above. This allows to recalibrate them along the measurement campaign and to improve  the correlation between quantum and classical accelerometers. In practice, these estimates are obtained using equations similar to \ref{eq:4} and \ref{eq:5}:

\begin{equation}
\hat{\theta}_{i+1}=\hat{\theta}_{i}+G_{\theta}(\hat{a}_{q,i}-\hat{a}_{c,i})\times X_{i} /(X_{i}^2+\sigma_X^2)
\end{equation}

for Algo I, and :

\begin{equation}
   \hat{\theta}_{i+1} = \hat{\theta}_i + G_{\theta}^{'} \times \frac{1}{k T^2} N_i \frac{D'_i}{(D'_i)^2 + \sigma_{D}'^2}
\end{equation}
with 
\begin{align}
    D'_i&=
    (\sin \hat{\phi}_{i-2}-\sin \hat{\phi}_{i-1} )\nonumber\\
    &\times (X_i \cos \hat{\phi}_i - X_{i-1}\cos \hat{\phi}_{i-1})\nonumber\\
    &- (\sin \hat{\phi}_i-\sin \hat{\phi}_{i-1} ) \nonumber\\
    &\times (X_{i-2} \cos \hat{\phi}_{i-2}- X_{i-1}\cos \hat{\phi}_{i-1}) 
\end{align}

for Algo II. $X$ stands for any of the four quantities $a_{x}, a_{y}, a'_{z}, a''_{z}$ derived from the measurements by the classical sensor, and $\hat{\theta}$ is the estimator for the corresponding sensitivity coefficient. An adapted gain is applied for each coefficient. 

Fig. \ref{fig:Figure05} displays the bias and the sensitivity coefficients when the coefficients from Table \ref{tab:parametres} are corrected by the algorithms. Some of the sensitivity coefficients determined with the algorithms exhibit significant differences with respect to the predetermined values (15 \% on average for $\eta_x$ and 6 \% on average for \textit{$\eta'$}). These variations could be due to distortions when installing the sensor onboard or to the different environmental conditions during the flight campaign, such as for example the operating temperature. Remarkably, the bias now features a more linear trend, with offsets related to half turns at the beginning and at the middle of the line suppressed.

Fig. \ref{fig:Figure06} compares the bias estimations when optimizing bias and scale factor only (Fig. \ref{fig:Figure06}.(a)), when optimizing all sensitivity coefficients, except the axis crossing coefficient along x (Fig. \ref{fig:Figure06}.(b)), and when optimizing all sensitivity coefficients (Fig. \ref{fig:Figure06}.(c)). The difference between Fig. \ref{fig:Figure06}.(b) and Fig. \ref{fig:Figure06}.(c) clearly demonstrates that the improvement in linearity is related to the better estimate of the alignment parameter $\eta_x$.

In addition to efficiently correct offsets in the bias estimation, we observe a significant improvement in the stability. Here, we will use a different figure of merit to characterize the algorithm efficiency in terms of stability since, by contrast with the simulations based on synthetic data, the Allan standard deviation of the bias estimates do not decrease as white noise, due to bias drift. Thus, we use instead as a criterion the standard deviation of the difference between the measured atomic acceleration, obtained via direct extraction, and its estimate by the classical sensor, corrected by one or the other of the algorithms. For Algo I, this standard deviation improves from $7.6\times10^{-5}$ m/s$^2$ when optimizing bias and scale factor only (Fig. \ref{fig:Figure06}.(a)) down to $5.2\times10^{-5}$ m/s$^2$, when optimizing all sensitivity coefficients together (Fig. \ref{fig:Figure06}.(c)). Performing tests with optimizing successively a single sensitivity coefficients showed that the improvement in the stability is dominated by the optimization of the filter parameter $\eta'$. Subtracting the contribution of the detection noise, which impacts the standard deviation by as much as $4.2\times10^{-5}$ m/s$^2$, the level of residual uncorrelated acceleration noise actually reduces by a factor 2 when optimizing the sensitivity coefficients, from 6.4 to $3.2\times10^{-5}$ m/s$^2$.

\section{Generalization to strapdown operation}

\begin{figure*}
    \centering
    \includegraphics[width=0.8\linewidth]{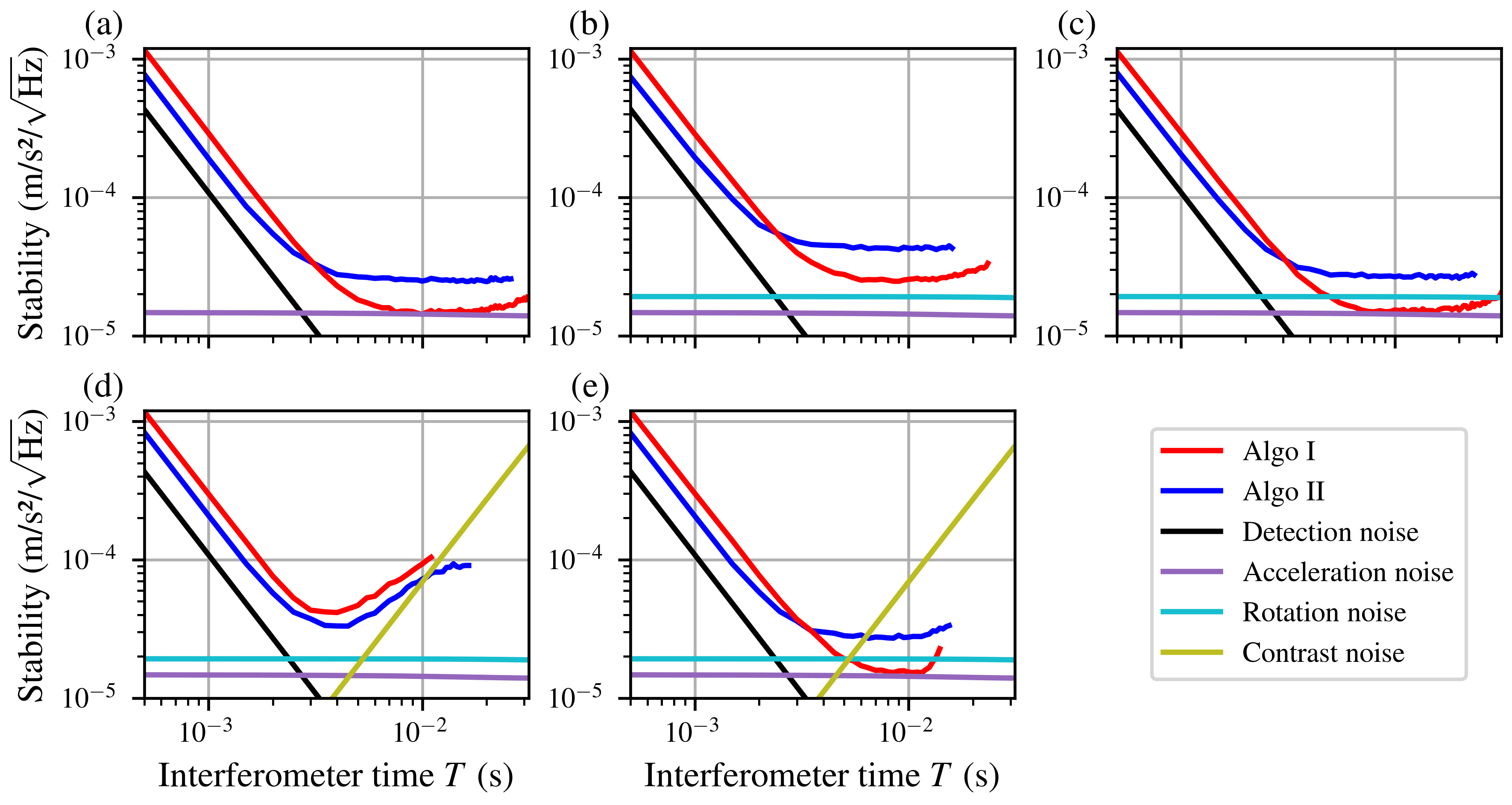}
    \caption{Stability of the bias estimates as a function of the interferometer time $T$ under different improved noise conditions and adaptive correction algorithms. (a) The QA is subject to a bias and a scale factor error, both estimated by the algorithms. (b) A rotational noise is added to the QA phase, but not corrected for. (c) The rotational noise is corrected for by the algorithms. (d) The contrast loss induced by rotations is taken into account in the QA model, but not in the algorithms. (e) This contrast loss is taken into account by the algorithms.}
    \label{fig:Figure07}
\end{figure*}

\begin{figure*}
    \centering
    \includegraphics[width=0.8\linewidth]{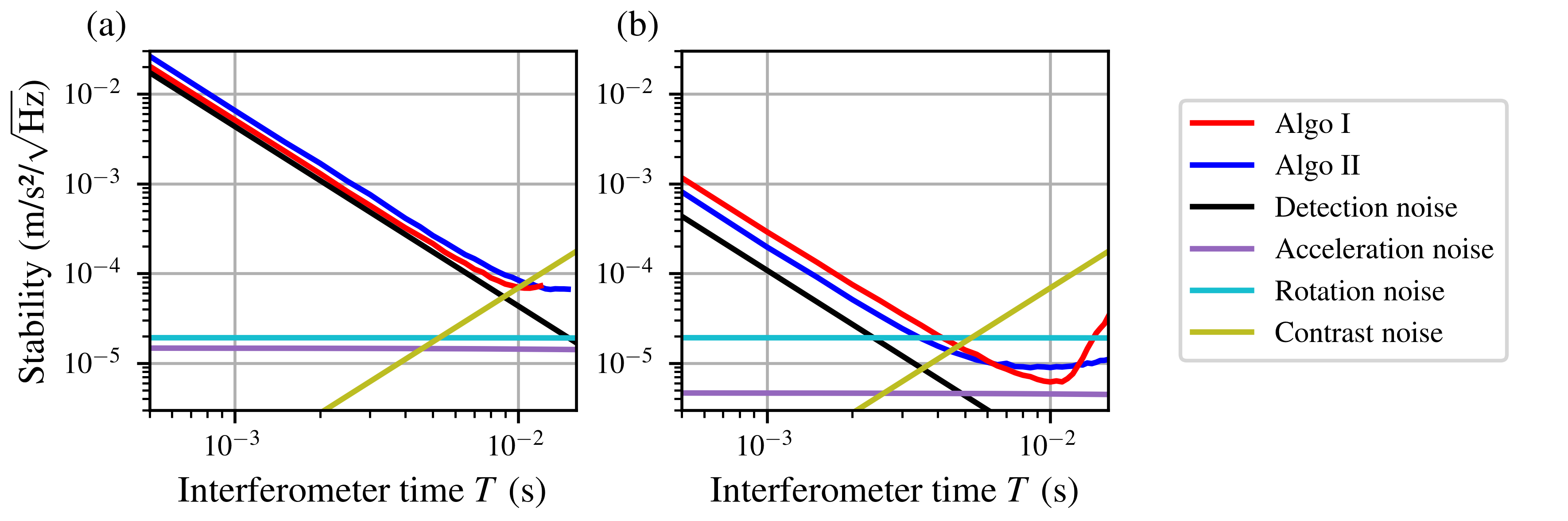}
    \caption{Stability of the bias estimates as a function of the interferometer time $T$ using the fully optimized algorithms for the Greenland campaign data. (a) typical conditions : The detection noise ($\sigma_P=0.016$) and the uncorrelated acceleration noise correspond to the GIRAFE instrument. (b) Improved conditions : The detection noise was reduced by a factor of about 30 and uncorrelated acceleration noise was reduced by a factor 3.}
    \label{fig:Figure08}
\end{figure*}

The algorithms can furthermore be generalized to account for additional phase contributions arising from rotations, which are negligible in the gyrostabilized mode of operation of GIRAFE, but would have an impact for strapdown measurements. Rotation induced acceleration terms are proportional to an internal parameter of the instrument such as the atomic initial velocity (for the QA). Those parameters can vary in time and have to be calibrated. They can thus be treated in the algorithm as sensitivity coefficients analogously to the ones in table \ref{tab:parametres}, provided that ancillary measurements of the rotation velocities and accelerations are available (using for instance classical gyroscopes). In this section, we highlight the potential use of the algorithms to retrieve the Coriolis acceleration bias, in a context where the contributions of Euler and centrifugal forces are assumed to be negligible.

Coriolis accelerations impacts a rotating QA in several ways. First, the interferometer phase contains two additional acceleration contributions\cite{Duan2020,Glick2024} : $2 v_{x0} \Omega_y$ and  $- 2v_{y0}\Omega_x$, with $v_{x0}$ and $v_{y0}$ the mean atomic velocities at the first interferometer pulse and $\Omega_x$ and $\Omega_y$ the angular velocities along the transverse directions. These velocity dependent terms lead to bias and noise related to fluctuations of the angular velocities and/or initial velocities of the atoms. Second, it leads to a loss of contrast for the interferometer related to the velocity spread of the atom source. The contrast then decreases exponentially with the interferometer duration following \cite{dArmagnacdeCastanet2024,Lan2012}:
\begin{equation}
\label{eq:ci}
C_i= C_0 \exp \left( - 2 \sigma_v^2  k_{eff}^2 (\Omega_{x,i}^2+\Omega_{y,i}^2) T^4 \right)
\end{equation}
where $\sigma_v$ is the standard deviation of the atomic velocity distribution.

This loss of contrast strongly impacts the performance, by on one hand decreasing the signal to noise ratio, effectively limiting the interferometer duration, and on the other hand inducing contrast fluctuations that cannot be distinguished from phase fluctuations by the algorithms. This contrast noise limits the stability of the bias estimation, with an impact on its Allan standard deviation at 1s that we calculated to be:
\begin{equation}
\label{eq:sigmac}
    \sigma_{C,1s} \approx 2\sqrt{3} k_{eff} T^2 \sigma_v^2 \sigma_\Omega^2 \sqrt{T_c}
\end{equation}
where $\sigma_\Omega$ is the standard deviation of temporal fluctuations of the angular velocity, and $T_c$ is the cycle time. 

We simulated hybrid measurement sessions in realistic strapdown conditions. For the acceleration and angular velocities, we used representative synthetic data drawn in independent normal distributions of standard deviations equal to those measured at a rate of 1 kHz by a strapdown Inertial Measurement Unit (IMU) during the Greenland campaign \cite{Jensen2025} (see table \ref{table:sigmaomega}). We set in the simulation a transverse initial velocity of the atom cloud of $v_0=2~\text{mm} \text{s}^{-1}$ and an atomic temperature of $2~\mu \text{K}$. When applicable, the Coriolis acceleration is integrated to the algorithms following the same procedure as in section \ref{sectioniv}: we add an estimate on both Coriolis terms, with the angular velocity measured by a classical gyrometer, and initial velocity components $v_{x0}$ and $v_{y0}$ as sensitivity factors to be estimated by the algorithms. 

\begin{table}[h]
    \centering
    \begin{tabular}{|l|c|}
        \hline
        \textbf{Parameter} & \textbf{Value} \\
        \hline
        $\sigma_a$     & $0.38~\text{m}.\text{s}^{-2}$ \\
        $\sigma_{\Omega x}$     & $5~\text{mrad}.\text{s}^{-1}$ \\
        $\sigma_{\Omega y}$     & $13~\text{mrad}.\text{s}^{-1}$ \\
        $\sigma_{\Omega z}$     & $11~\text{mrad}.\text{s}^{-1}$ \\
        \hline
    \end{tabular}
    \caption{Standard deviations of navigation parameters obtained from measurements during the Greenland campaign, used in the simulation.}
    \label{table:sigmaomega}
\end{table}

Fig. \ref{fig:Figure07} displays the stability of the bias estimates by both algorithms, as a function of the interferometer duration $T$. Five different cases are considered, with in all cases a bias of $2\times10^{-5}~\text{m}~\text{s}^{-2}$ and initial errors of $0.01$ \% in the scale factor and of $2~\text{mm} \text{s}^{-1}$ in the initial velocity. If the algorithms fail to converge, i.e. when the integration time is large, stability is undefined and not shown on the graphs.

 Fig.\ref{fig:Figure07} (a) shows the results obtained without rotational contributions, with the algorithms evaluating and correcting only the acceleration bias and scale factor of the CA. The bias stability is limited by detection noise at low $T$ and by uncorrelated variations in acceleration at higher $T$. Note that for $T \gtrsim 30$ ms, the two algorithms do not converge, because the level of uncorrelated acceleration then lead to flaws in the error estimation. The black line displays the limit to the bias stability due to detection noise, which can be calculated analytically as: 
 \begin{equation}
 \sigma_{D,1s} \approx \frac{2\sqrt{2}}{C k_{eff} T^2 }\sqrt{Tc} \ \sigma_p
\end{equation}

 Fig.\ref{fig:Figure07} (b) illustrates the degradation of the performances due to the presence of rotations. Here, we add rotational contributions to the interferometer phase, and thus to the acceleration measured by the QA, but do not correct from them. We observe a degradation of the Allan standard deviation for $T \gtrsim 2-3$ ms, caused by fluctuations of Coriolis accelerations, which set the limit displayed as a cyan line. 

In Fig.\ref{fig:Figure07} (c), the algorithms are set up to best estimate $v_0$ and correct the QA acceleration measurements from the contribution of Coriolis accelerations, bringing back the Allan standard deviations to the levels obtained in the first case. Note that here, the loss of contrast induced by the rotations is not taken into account, corresponding to an ideal case where the atom temperature would be null. 

Fig.\ref{fig:Figure07} (d) accounts for the contrast loss with a realistic atomic temperature of $2~\mu$K, corresponding to a rms atomic velocity of $\sigma_v\sim 1$ cm s$^{-1}$. The loss of contrast is taken into account in the generation of the transition probabilities, but not in the derivation by the algorithms of the difference of the accelerations measured by the two sensors. This leads to a significant decrease of the stability for both algorithms, with an optimum reached at around $T = 4 ms$. The yellow line displays the limit to the bias stability due to contrast noise given by equation (\ref{eq:sigmac}).

In Fig.\ref{fig:Figure07} (e), we adapt the algorithms to take into account the fluctuating contrast in the derivation of the difference of the accelerations measured by the two sensors, assuming the temperature is exactly known and using equation (\ref{eq:ci}). This improves notably the stability for long interferometer durations, efficiently suppressing the detrimental effects associated to the strapdown operation. The optimal stabilities for both Algo I and Algo II are indeed now comparable to the stabilities in case a), which would correspond to a gyrostabilized mode of operation with an improved detection noise. 

Finally, figure \ref{fig:Figure08} displays the performances of the full algorithms, as implemented in  Fig.\ref{fig:Figure07} (e), in the case a) where the level of detection and uncorrelated acceleration noises correspond to the GIRAFE sensor, and in the case b) where detection noise is reduced by a factor of 30 and the level of correlation between the QA and the CA is improved by a factor of 3. We show here a potential improvement by one order of magnitude on the stability, highlighting the possibility to improve the quality of the measurements even in strapdown mode. 

\section{Conclusion}

In this paper, we have compared two different hybridization algorithms, one performing slightly better in terms of terms of noise, the second one showing better robustness with respect to errors in parameter estimation. Both allow to efficiently retrieve the bias and the scale factor of classical acceleration sensors in onboard operation on a gyrostabilized platform and further improve the hybridization by determining other key accelerometer parameters. We have shown how additional signal contributions arising from rotational accelerations can also be accounted by deepening the hybridization to rotation sensors. Further improvements of the algorithms could include dynamic correction of the contrast noise related term, which unlike what has been done in this paper does not have a linear relation to the measured inertial quantity. These optimizations extend the range of operation of quantum sensors to more aggressive operating conditions, and opens in particular perspectives for airborne gravity measurement campaigns in strapdown mode with improved performances. 


\section*{Acknowledgments}
\paragraph*{Funding:}
This work was supported in part by a government grant managed by the Agence Nationale de la Recherche under the Plan France 2030 with the reference "ANR-22-PETQ-0005". This project has received funding from the European Defence Fund (EDF) under grant agreement 101103417 EDF-2021-DIS-RDIS-ADEQUADE, funded by the European Union.

\paragraph*{Competing interests:}
The authors declare that they have no competing interests.
\paragraph*{Data and materials availability:}
All data needed to evaluate the conclusions made in this work are present in the text or supplementary material and can be made available to the reader upon reasonable request of the authors.

\paragraph*{Note:}
Prior to publication, please use: 
``The following article has been submitted to/accepted by AVS Quantum Science. 
After it is published, it will be found at \url{https://aqs.peerx-press.org/ms_files/aqs/2025/09/30/02005892/00/2005892_0_art_file_25647548_t3yvdy.pdf} (If clicking the link gives a ‘Forbidden’ error, please copy and paste it directly into your browser.)

\paragraph*{License:}
This preprint is made available under the Creative Commons Attribution-NonCommercial 4.0 International (CC BY-NC-SA 4.0) License.\url{https://creativecommons.org/licenses/by-nc/4.0/}
\nocite{*}
\bibliography{biblio}

\end{document}